\documentstyle[preprint,prc,aps,eqsecnum,epsf]{revtex}
\tighten
\begin{document}
\draft

\title{Weak Transitions in $A$=6 and 7 Nuclei}
\author{R.\ Schiavilla}
\address{Jefferson Lab, Newport News, Virginia 23606 \\
         and \\
         Department of Physics, Old Dominion University, Norfolk, Virginia 23529}
\author{R.B.\ Wiringa}
\address{Physics Division, Argonne National Laboratory, Argonne, Illinois 60439}
\date{\today}
\maketitle

\begin{abstract}
The $^6$He beta decay and $^7$Be electron capture
processes are studied using variational Monte Carlo wave
functions, derived from a realistic Hamiltonian
consisting of the Argonne $v_{18}$ two-nucleon and Urbana-IX
three-nucleon interactions.  The model for the nuclear weak axial current
includes one- and two-body operators with the strength of the leading
two-body term---associated with $\Delta$-isobar excitation
of the nucleon---adjusted to reproduce the Gamow-Teller matrix
element in tritium $\beta$-decay.  The measured half-life of $^6$He
is under-predicted by theory by $\simeq$ 8\%, while that of $^7$Be
for decay into the ground and first excited states of $^7$Li is
over-predicted by $\simeq$ 9\%.  However, the experimentally
known branching ratio for these latter processes is in good agreement
with the calculated value.  Two-body axial current contributions
lead to a $\simeq$ 1.7\% (4.4\%) increase in the value of
the  Gamow-Teller matrix element of $^6$He ($^7$Be), obtained with
one-body currents only, and slightly worsen (appreciably improve)
the agreement between the calculated and measured half-life.
Corrections due to retardation effects associated with the finite
lepton momentum transfers involved in the decays, as well as contributions 
of suppressed transitions induced by the weak vector charge and
axial current operators, have also been calculated and found to be
negligible.  
\end{abstract}

\section{Introduction}
\label{sec:intro}

The present study deals with weak transitions in the $A$=6 and 7
systems within the context of a fully microscopic approach to nuclear
structure and dynamics, in which nucleons interact among themselves
via realistic two- and three-body potentials, and with external
electroweak probes via realistic currents consisting
of one- and many-body components.  To the best of our knowledge,
calculations of the superallowed $^6$He beta-($\beta$-) decay and $^7$Be
electron-($\epsilon$-) capture processes have relied in the past on relatively
simple shell-model or two- and three-cluster wave functions.
The shell model calculations have typically failed to
reproduce the measured Gamow-Teller matrix elements
governing these weak transitions, unless use was made of an effective
one-body Gamow-Teller operator, in which the nucleon's axial coupling constant
$g_A$ is quenched with respect to its free value---for a recent
summary of a shell-model analysis of $\beta$-decay rates in $A \leq$ 18
nuclei, see Ref.~\cite{Chou93}.

More phenomenologically successful models have been based on
$\alpha NN$ (for $A$=6) or $\alpha$-$t$ and $\alpha$-$\tau$
(for $A$=7) clusterization, and have used either Faddeev
techniques with a separable representation of the $N$$N$ and
$\alpha$$N$ potentials~\cite{Parke78} or the resonating-group
method~\cite{Walliser83}.  However, while these studies do provide
useful insights into the structure of the $A$=6 and 7 nuclei, their
connection with the underlying two- (and three-) nucleon dynamics
is rather tenuous.  For example, it is unclear whether the required
quenching of $g_A$ in the shell-model calculations reflects deficiencies
in the associated wave functions---a lack of correlations
or limitations in the model space---and/or in the model for the
axial current operator---which typically ignores many-body terms---or,
rather, a true modification of the nucleon axial coupling in medium. 

In this work, we use variational Monte Carlo (VMC) wave 
functions~\cite{W91,PPCPW97,NWS01,N01},
derived from a realistic Hamiltonian consisting of the Argonne $v_{18}$
two-nucleon~\cite{Wiringa95} and Urbana-IX three-nucleon~\cite{Pudliner95}
interactions.  The VMC wave functions provide a reasonable description
of the energy spectra of low-lying states in $A$=6--8 
nuclei~\cite{PPCPW97,Wiringa00}, and of elastic and inelastic electromagnetic 
form factors and radiative widths of $^6$Li states~\cite{Wiringa98}.  

The model for the nuclear weak vector and axial-vector
currents is that developed in Refs.\cite{Schiavilla98,Marcucci01},
consisting of one- and two-body terms.
The weak vector charge and current operators are
constructed from their isovector electromagnetic
counter-parts~\cite{Schiavilla89,Schiavilla90}, in accordance with
the conserved-vector-current (CVC) hypothesis.
The leading two-body term in the axial current
is due to $\Delta$-isobar excitation, while the leading
two-body axial charge operator is associated with the long-range
pion-exchange term~\cite{Kubodera78}, required by low-energy
theorems and the partially-conserved-axial-current relation.
The largest model dependence is in the weak axial current.  To minimize it,
the $N \rightarrow \Delta$ transition axial coupling constant
has been adjusted to reproduce the experimental value
of the Gamow-Teller matrix element in tritium $\beta$-decay in
an essentially exact calculation, using correlated-hyperspherical-harmonics
wave functions, derived from the same
Hamiltonian adopted here~\cite{Marcucci01a}.  

This manuscript falls into seven sections.  In
Sec.~\ref{sec:rate} explicit expressions for the
$\beta$-decay and $\epsilon$-capture rates are derived
in terms of reduced matrix elements of multipole
operators of the nuclear weak current, whose
model is succinctly described in Sec.~\ref{sec:cur}.
The $A$=6 and 7 nuclei VMC wave functions are reviewed
in Sec.~\ref{sec:waves}, while predictions for
the $^6$He $\beta$-decay and $^7$Be $\epsilon$-capture rates
are presented and discussed in Sec.~\ref{sec:res}.
Our conclusions are summarized in Sec.~\ref{sec:concl}.
\section{Transition Rates}
\label{sec:rate}

Nuclear $\beta$-decay is induced by the weak-interaction
Hamiltonian~\cite{Walecka95}

\begin{equation}
H_{W}=\frac{G_{V}}{\sqrt{2}}\int {\rm d}{\bf{x}}\,{\rm{e}}^{-{\rm{i}}
({\bf{p}}_{e}+{\bf{p}}_{\nu})\cdot{\bf{x}}}\,l_{\sigma}\,j^{\sigma}({\bf{x}})
\ , \label{eq:hw}
\end{equation}
where $G_{V}$ is the Fermi coupling constant
($G_{V}$=1.14939 10$^{-5}$ GeV$^{-2}$~\cite{Hardy90}),
$l_{\sigma}$ is the leptonic weak current

\begin{equation}
l_{\sigma} = \overline{u}_e\gamma_{\sigma}(1-\gamma_5)v_{\nu} \ ,
\label{eq:lmu}
\end{equation}
and $j^{\sigma}({\bf{x}})$ is the hadronic weak current density.
The electron and (electron) antineutrino momenta and spinors are
denoted, respectively, by ${\bf{p}}_{e}$ and ${\bf{p}}_{\nu}$, and
$u_{e}$ and ${v}_{\nu}$ ($e^-$ emission is being discussed
here). The Bjorken and Drell~\cite{Bjorken64}
conventions are used for the metric tensor $g^{\sigma\tau}$ and
$\gamma$-matrices.  However, the spinors are normalized as
$u_{e}^{\dag}u_{e}={v}_{\nu}^{\dag}{v}_{\nu}=1$.  The amplitude
for the process $^AZ\rightarrow ^A\!(Z+1)e^- \overline{\nu}_e$ is then given by

\begin{equation}
\langle f|H_{W}|i\rangle=
\frac{G_{V}}{\sqrt{2}}\,l^{\sigma} \langle -{\bf{q}}; ^A\!(Z+1),J_f\,M_f|
j_{\sigma}^{\dag}({\bf{q}})|^AZ,J_i\, M_i\rangle \ ,
\label{eq:tra}
\end{equation}
where ${\bf{q}}={\bf{p}}_{e}+{\bf{p}}_{\nu}$,
$|^AZ,J_i\, M_i\rangle$ and $|^A(Z+1),J_f\, M_f\rangle$
represent the initial and final nuclear states, the
latter recoiling with momentum $-{\bf q}$, with spins
$J_i$ and $J_f$ and spin projections $M_i$ and $M_f$,
respectively, and

\begin{equation}
j^{\sigma}({\bf{q}})=\int {\rm d}{\bf{x}}\,
{\rm{e}}^{{\rm{i}}{\bf{q}}\cdot{\bf{x}}} \,j^{\sigma}({\bf{x}})\equiv
(\rho({\bf{q}}),{\bf{j}}({\bf{q}})) \ .
\label{eq:jmuq}
\end{equation}

Standard techniques~\cite{Marcucci01,Walecka95} are
now used to carry out the multipole expansion
of the weak charge, $\rho({\bf q})$, and current,
${\bf j}({\bf q})$, operators in a reference frame
in which the $\hat{\bf z}$-axis defines the
spin-quantization axis, and the direction $\hat{\bf q}$
is specified by the angles $\theta$ and $\phi$:

\begin{eqnarray}
\langle J_f\,M_f| \rho^\dagger({\bf q}) |J_i\, M_i\rangle
&=&\sqrt{4\pi} \sum_{l\geq 0} X^l_0 \, C_l(q)
\ , \label{eq:c} \\
\langle J_f\,M_f| \,{\hat{\bf{e}}}^{*}_{q0}\cdot
{\bf{j}}^{\dag}({\bf{q}})\, |J_i\, M_i\rangle
&=&\sqrt{4\pi} \sum_{l\geq 0} X^l_0 \, L_l(q)
\ , \label{eq:l} \\
\langle J_f\,M_f| \,{\hat{\bf{e}}}^{*}_{q\pm 1}\cdot
{\bf{j}}^{\dag}({\bf{q}})\, |J_i\, M_i\rangle
&=&-\sqrt{2\pi} \sum_{l\geq 1} X^l_{\mp1} \,\left[\pm M_l(q)+E_l(q) \right]
\ , \label{eq:em}
\end{eqnarray}
where the dependence on the direction $\hat{\bf q}$, and
on the initial and final spins and spin projections is
contained in the functions $X^l_\lambda$, with $\lambda$=$0,\pm 1$,
defined as

\begin{equation}
X_\lambda^l(\hat{\bf q};J_f\, M_f,J_i\, M_i) \equiv (-{\rm i})^l\, 
\Bigg(\frac{2\, l+1}{2\, J_f+1}\Bigg)^{1/2}\, D^l_{l_z,\lambda}(-\phi,-\theta,0) \,
\langle J_i\, M_i, l\, l_z | J_f\, M_f\rangle \ ,
\label{eq:xfn}
\end{equation}
with $l_z$=$M_f-M_i$.  In Eqs.~(\ref{eq:c})--(\ref{eq:em}) the
$q$-dependent coefficients are the reduced matrix elements (RME's)
of the Coulomb $(C)$, longitudinal $(L)$, transverse electric $(E)$
and transverse magnetic $(M)$ multipole operators, as given 
in Refs.~\cite{Marcucci01,Walecka95}, and the vectors $\hat{\bf e}_\lambda$
denote the linear combinations

\begin{eqnarray}
\hat{{\bf{e}}}_{q0}&\equiv&
{\hat{\bf e}}_{q3} \ , \label{eq:eq0} \\
\hat{{\bf{e}}}_{q\pm{1}}&\equiv&
\mp\frac{1}{\sqrt{2}}(\hat{{\bf{e}}}_{q1}
\pm\,{\rm{i}}\,\hat{{\bf{e}}}_{q2}) \ , \label{eq:eqpm1}
\end{eqnarray}
with $\hat{\bf{e}}_{q3}=\hat{\bf q}$,
$\hat{\bf{e}}_{q2}=\hat{\bf z} \times {\bf q}/|\hat{\bf z} \times {\bf q}|$,
$\hat{\bf{e}}_{q1}= \hat{\bf{e}}_{q2} \times \hat{\bf{e}}_{q3}$.
Since the weak charge and current operators have
scalar/polar-vector $(V)$ and pseudoscalar/axial-vector $(A)$
components, each multipole consists of the sum of $V$ and $A$ terms:

\begin{equation}
T_{ll_z}=T_{ll_z}(V)+T_{ll_z}(A) \ ,
\end{equation}
where $T$=$C$, $L$, $E$, and $M$, and
the parity of $l$th-pole $V$-operators is opposite of that of
$l$th-pole $A$-operators.  The parity of Coulomb, longitudinal, and
electric $l$th-pole $V$-operators is $(-)^l$, while that of magnetic
$l$th-pole $V$-operators is $(-)^{l+1}$.
Finally, in Eq.~(\ref{eq:xfn}) the $D^l_{l_z,l_z^\prime}$
are rotation matrices in the standard notation of Ref.~\cite{Edmonds57}.

The rate for nuclear beta decay is then obtained from 

\begin{equation}
{\rm d}\Gamma^\beta_{fi}=2\pi\, \delta(E_i-E_f) \frac{1}{2\, J_i+1}
\sum_{M_i M_f} \sum_{s_e s_\nu}
\mid\!\langle f|H_{W}|i\rangle\!\mid^2 \frac{{\rm d}{\bf p}_e}{(2\pi)^3}
\frac{{\rm d}{\bf p}_\nu}{(2\pi)^3} \ ,
\label{eq:rtt}
\end{equation}
where $E_i$=$m_i$ is the rest mass of the initial
nucleus $^AZ$, $E_f$=$p_\nu+\sqrt{p_e^2+m_e^2}+\sqrt{q^2+m_f^2}$
is the energy of the final state, $m_e$ and $m_f$ being the rest masses
of the electron and final nucleus $^A(Z+1)$.  Carrying out the spin
sums leads to~\cite{Marcucci01,Walecka95}

\begin{eqnarray}
\frac{1}{2\, J_i+1}\!\sum_{M_i M_f} \sum_{s_e s_\nu}
\mid\!\langle f|H_{W}|i\rangle\!\mid^2&=&G_V^2\,\frac{4\pi}{2\, J_i+1}\Bigg[
(1+{\bf v}_e\!\cdot\! \hat{\bf v}_\nu) \sum_{l\geq 0} |C_l(q)|^2 \nonumber \\
&+&(1-{\bf v}_e \!\cdot\! \hat{\bf v}_\nu+2\,{\bf v}_e\!\cdot\! \hat{\bf q}
\,{\bf v}_\nu\!\cdot\! \hat{\bf q})\sum_{l\geq 0} |L_l(q)|^2 \nonumber \\
&-&2\, \hat{\bf q}\!\cdot\!({\bf v}_e+\hat{\bf v}_\nu)\, 
\sum_{l\geq 0} {\rm Re}\left[C_l(q) L_l^*(q)\right] \nonumber \\
&+&\!(1-{\bf v}_e\!\cdot\! \hat{\bf q}\,
{\bf v}_\nu\!\cdot\! \hat{\bf q})\sum_{l\geq 1}
\left[ |M_l(q)|^2+|E_l(q)|^2 \right] \nonumber \\
&-&2\, \hat{\bf q}\!\cdot\!({\bf v}_e-\hat{\bf v}_\nu)\sum_{l\geq 1}
{\rm Re}\left[ M_l(q)E_l^*(q)\right]\Bigg] \ ,
\label{eq:rate}
\end{eqnarray}
where ${\bf v}_e$=${\bf p}_e/\sqrt{p_e^2+m_e^2}$ and
$\hat{\bf v}_\nu$=$\hat {\bf p}_\nu$ are the velocities
of the electron and neutrino, respectively.  Note that angular
momentum and parity selection rules restrict the number of
non-vanishing $l$-multipoles contributing to the transition.

A few comments are now in order.  Firstly, the expression
above is valid for $e^-$ emission, however, the rate for
$e^+$ emission corresponding to a transition
$^AZ \rightarrow\, ^A(Z-1) e^+\nu_e$ has precisely
the same form, but for the sign in the last term of
Eq.~(\ref{eq:rate}), $+2\,\hat{\bf q}\!\cdot\!({\bf v}_e-\hat{\bf v}_\nu) \cdots\,\,$.
Secondly, for allowed transitions with $|J_i-J_f|=\pm 1,0$
and $\pi_i\, \pi_f$=1 the above
rate is easily shown to reduce to the familiar expression
in terms of Fermi [F $\propto C_0(q=0;V)$] and Gamow-Teller
[GT $\propto E_1(q=0;A)$=$\sqrt{2}\, L_1(q=0;A)$] matrix elements, see Sec.~\ref{sec:cal}
and Ref.~\cite{Walecka95}.  Finally, a more realistic treatment---like
that outlined in Ref.~\cite{Schopper66}, for instance---takes into
account the distortion of the outgoing $e^\pm$ wave function
in the Coulomb field of the residual nuclear system.
The simple approximation of multiplying
the right-hand-side of Eq.~(\ref{eq:rate}) by the ratio
of the charged lepton density at the nuclear radius to
the density at infinity has been deemed, however, to suffice for
our purposes here (see Sec.~\ref{sec:cal}).  One of the
objectives of the present study is to estimate,
in an allowed decay such as the $^6{\rm He}
\rightarrow ^6{\rm Li}\, e^-\, \overline{\nu}_e$
under consideration [with ($J_i^{\pi_i};T_i$)=(0$^+$;1)
and ($J_f^{\pi_f};T_f$)=(1$^+$;0)], the size of corrections
associated with i) retardation effects due to the
finite lepton momentum transfer in the decay, and ii) transitions
other than those of F [$C_0(q;V)$] and/or GT
[$L_1(q;A)$ and $E_1(q;A)$] type.  For example,
a naive analysis of the 6-body decay above would ignore
the momentum transfer dependence---it is in the
range $0 \leq q \leq 4$ MeV/c---as well as the
contributions arising from transitions induced by
the axial charge and vector current operators via $C_1(q;A)$
and $M_1(q;V)$ RME's, respectively.  Of course,
available tabulations of F and GT matrix elements
extracted from experiment~\cite{Raman78} do attempt to
estimate these corrections (as well as those
due to electron screening, the finite extent of the
nuclear charge distribution, etc.).  The latter 
are typically obtained, however, within the context
of a shell-model description of the nuclear wave functions. 
Thus, it is interesting to re-examine the issue above
within the present framework using more realistic
wave functions.

The $e^-$ capture (so-called $\epsilon$-capture)
process is governed by the same weak-interaction
Hamiltonian in Eq.~(\ref{eq:hw}).  However, the
lepton weak-current density is now given by

\begin{equation}
l_{\sigma}({\bf x})={\rm e}^{-{\rm i} {\bf p}_\nu \cdot {\bf x} } \,
\overline{u}_\nu \gamma_\sigma (1-\gamma_5)
\langle{\rm Atom}_f\mid \psi_e({\bf x})\mid{\rm Atom}_i\rangle \ ,
\label{eq:lepc}
\end{equation}
where $\psi_e({\bf x})$ is the electron field operator,
and $\mid{\rm Atom}_i\rangle$ and $\mid{\rm Atom}_f\rangle$
are the initial and final atomic states, respectively.
A realistic description of nuclear electron capture
requires, therefore, a careful treatment of the atomic
physics aspects of the process~\cite{Bambynek77},
such as those relating to atomic wave-function overlaps,
exchange contributions, and electron-correlation effects.
We defer to Sec.~\ref{sec:cal} for a discussion of some
of these issues in the context of the $^7$Be $\epsilon$-capture
of interest here.  In this section, however, we simply
approximate

\begin{equation}
\langle{\rm Atom}_f\mid \psi_e({\bf x})\mid{\rm Atom}_i\rangle
\simeq \frac{R_{1s}(x)}{\sqrt{4\pi}} \chi(s_e) \equiv
\frac{R_{1s}(x)}{\sqrt{4\pi}} u({\bf p}_e,s_e)
\qquad p_e \rightarrow 0 \ ,
\label{eq:atom}
\end{equation}
ignoring atomic many-body effects and relativistic
corrections---this latter approximation is reasonably
justified for low $Z$, since the $e^-$ velocity
is $\simeq Z\alpha \ll 1$.  In Eq.~(\ref{eq:atom}), $R_{1s}(x)$
is the $1s$ radial solution of the Schr\"odinger
equation, and the two-component spin state $\chi(s_e)$
has been conveniently replaced by the four-component
spinor $u({\bf p}_e,s_e)$ in the limit $p_e \rightarrow 0$, which
allows us to use standard techniques to carry out
the spin sum over $s_e$ at a later stage.  The transition
amplitude reads

\begin{equation}
\langle f|H_{W}|i\rangle=
\frac{G_{V}}{\sqrt{8\pi}}\,R_{1s}(0)\,
\tilde{l}^{\sigma}\, \langle -{\bf p}_\nu; ^A\!(Z-1),J_f\,M_f|
j_{\sigma}^{\dag}({\bf p}_\nu)|^AZ,J_i\, M_i\rangle \ ,
\label{eq:trae}
\end{equation}
where $\tilde{l}_\sigma\equiv \overline{u}_\nu\gamma_\sigma(1-\gamma_5)u_e$,
and the $e^-$ radial wave function has been factored out from the matrix element
of $j^\sigma({\bf p}_\nu)$ by approximating it with its value at the origin.
The resulting differential rate is then written as~\cite{Marcucci01a,Walecka95}

\begin{equation}
{\rm d}\Gamma^{\epsilon}_{fi}=2\pi\, \delta(E_i-E_f)\, \frac{1}{2\, J_i+1}
\sum_{M_i M_f} \sum_{s_e s_\nu}
\mid\!\langle f|H_{W}|i\rangle\!\mid^2
\frac{{\rm d}{\bf p}_\nu}{(2\pi)^3} \ ,
\label{eq:rtte}
\end{equation}
where the initial and final energies
are now given by $E_i$=$m_i+m_e$ and
$E_f$=$p_\nu+\sqrt{p_\nu^2+m_f^2}$,
respectively.  Note that atomic binding energy contributions
have been neglected, since they are of the order
$(Z\alpha)^2\, m_e/2$.  The square of the amplitude
summed over the spins can be obtained{\it, mutatis mutandis},
from that corresponding to $e^+$ emission discussed above in
the limit $p_e$=0 (${\bf q}$=${\bf p}_\nu$):

\begin{eqnarray}
\frac{1}{( 2\, J_i+1)}\!\sum_{M_i M_f} \sum_{s_e s_\nu}
\mid\!\langle f|H_{W}|i\rangle\!\mid^2&=&G_V^2\, 
\frac{|R_{1s}(0)|^2}{4\pi}\, \frac{4\pi}{(2\, J_i+1)}\, \Bigg[
 \sum_{l\geq 0} |C_l(p_\nu)-L_l(p_\nu)|^2 \nonumber \\
&+&\sum_{l\geq 1} |M_l(p_\nu)-E_l(p_\nu)|^2 \Bigg] \ .
\label{eq:ratee}
\end{eqnarray}
\section{Wave Functions}
\label{sec:waves}

The VMC wave function, $\Psi_T(J^\pi;T)$, for a given nucleus, is constructed 
from products of two- and three-body correlation operators acting on an 
antisymmetric single-particle state with the appropriate quantum numbers.
The correlation operators are designed to reflect the influence of the 
interactions at short distances, while appropriate boundary conditions
are imposed at long range~\cite{W91,PPCPW97,NWS01,N01}.
The $\Psi_T(J^\pi;T)$ has embedded variational parameters
that are adjusted to minimize the expectation value
\begin{equation}
   E_T = \frac{\langle \Psi_T | H | \Psi_T \rangle}
              {\langle \Psi_T   |   \Psi_T \rangle} \geq E_0 \ ,
\label{eq:expect}
\end{equation}
which is evaluated by Metropolis Monte Carlo integration.

A good variational trial function has the form
\begin{equation}
     |\Psi_T\rangle = \left[1 + \sum_{i<j<k}\tilde{U}^{TNI}_{ijk} \right]
                      \left[ {\cal S}\prod_{i<j}(1+U_{ij}) \right]
                      |\Psi_J\rangle \ .
\label{eq:psit}
\end{equation}
The Jastrow wave function, $\Psi_J$, is fully antisymmetric and has the
$(J^\pi;T)$ quantum numbers of the state of interest.
For the $s$-shell nuclei we use the simple form
\begin{equation}
     |\Psi_J\rangle = \left[ \prod_{i<j<k}f^c_{ijk} \right]
                      \left[ \prod_{i<j}f_c(r_{ij}) \right]
                     |\Phi_A(JMTT_{3})\rangle \ .
\label{eq:jastrow}
\end{equation}
Here $f_c(r_{ij})$ and $f^c_{ijk}$ are central two- and three-body correlation
functions and $\Phi_A$ is a Slater determinant in spin-isospin space, e.g.,
for the $\alpha$-particle,
\begin{equation}
     |\Phi_{4}(0 0 0 0) \rangle
        = {\cal A} |p\uparrow p\downarrow n\uparrow n\downarrow \rangle \ .
\end{equation}
The $U_{ij}$ and $\tilde{U}^{TNI}_{ijk}$ are noncommuting two- and 
three-nucleon correlation operators, and ${\cal S}$
indicates a symmetric sum over all possible orders.
The $U_{ij}$ includes spin, isospin, and tensor terms:
\begin{equation}
     U_{ij} = \sum_{p=2,6} u_p(r_{ij}) O^p_{ij} \ ,
\end{equation}
where the $O^{p=1,6}_{ij}=[1, {\bf\sigma}_{i}\cdot{\bf\sigma}_{j}, S_{ij}]
\otimes[1,{\bf\tau}_{i}\cdot{\bf\tau}_{j}]$ are the main static operators 
that appear in the $N\!N$ potential.
The $f_c(r)$ and $u_p(r)$ functions are generated by the solution of a
set of coupled differential equations which contain a number of variational 
parameters~\cite{W91}.
The $\tilde{U}^{TNI}_{ijk}$ has the spin-isospin structure of the dominant
parts of the $N\!N\!N$ interaction as suggested by perturbation theory.

The optimal $U_{ij}$ and $\tilde{U}^{TNI}_{ij;k}$ do not change significantly
from nucleus to nucleus, but $\Psi_J$ does.
For the $p$-shell nuclei, $\Psi_J$ includes a one-body part that consists of
4 nucleons in an $\alpha$-like core and $(A-4)$ nucleons in $p$-shell orbitals.
We use $LS$ coupling to obtain the desired $JM$ value,
as suggested by standard shell-model studies~\cite{CK65}.
We also need to sum over different spatial symmetries $[n]$ of the angular
momentum coupling of the $p$-shell nucleons~\cite{BM69}.
The one-body parts are multiplied by products of central pair
and triplet correlation functions, which depend upon the shells ($s$ or $p$)
occupied by the particles and on the $LS[n]$ coupling:
\begin{eqnarray}
  |\Psi_{J}\rangle &=& {\cal A} \left\{\right.
     \prod_{i<j<k} f^{c}_{ijk}
     \prod_{i<j \leq 4}f_{ss}(r_{ij})
     \prod_{k \leq 4<l \leq A} f_{sp}(r_{kl})  \nonumber\\
                   && \sum_{LS[n]}
     \Big( \beta_{LS[n]} \prod_{4<l<m \leq A} f^{LS[n]}_{pp}(r_{lm})
    |\Phi_{A}(LS[n]JMTT_{3})_{1234:5\ldots A}\rangle \Big) \left.\right\} \ .
\label{eq:jastrow-p}
\end{eqnarray}
The operator ${\cal A}$ indicates an antisymmetric sum over all possible
partitions into 4 $s$-shell and $(A-4)$ $p$-shell particles.
The pair correlation for both particles within the $s$-shell, $f_{ss}$,
is set to the $f_c$ of the $\alpha$-particle.
The $f_{sp}$ is similar to $f_{ss}$ at short range, but with a
long-range tail going to a constant $\approx 1$, while the $f^{LS[n]}_{pp}$ 
is allowed to depend on the particular single-particle channel.

The $LS[n]$ components of the single-particle wave function are given by:
\begin{eqnarray}
 &&  |\Phi_{A}(LS[n]JMTT_{3})_{1234:5\ldots A}\rangle =
     |\Phi_{4}(0 0 0 0)_{1234}\rangle \prod_{4 < l\leq A}
     \phi^{LS[n]}_{p}(R_{\alpha l}) \\
 &&  \left\{ [ \prod_{4 < l\leq A} Y_{1m_l}(\Omega_{\alpha l}) ]_{LM_L[n]}
     \times [ \prod_{4 < l\leq A} \chi_{l}(\case{1}{2}m_s) ]_{SM_S}
     \right\}_{JM}
     \times [ \prod_{4 < l\leq A} \nu_{l}(\case{1}{2}t_3) ]_{TT_3}\rangle
     \nonumber \ .
\end{eqnarray}
The $\phi^{LS}_{p}(R_{\alpha l})$ are $p$-wave solutions of a particle
in an effective $\alpha$-$N$ potential that has Woods-Saxon and Coulomb parts.
They are functions of the distance between the center of mass
of the $\alpha$ core and nucleon $l$, and may vary with $LS[n]$.
The wave function is translationally invariant, so there is no
spurious center of mass motion.

Two different types of $\Psi_J$ have been constructed in recent VMC 
calculations of light $p$-shell nuclei: an original shell-model kind of trial 
function~\cite{PPCPW97} which we will call Type I, and a cluster-cluster
kind of trial function~\cite{NWS01,N01} which we will call Type II.
In Type I trial functions, the $\phi^{LS}_{p}(r)$ has an exponential decay
at long range, with the depth, range, and surface thickness of the 
Woods-Saxon potential serving as variational parameters.
The $f_{sp}$ goes to a constant somewhat less than unity, while $f^{LS[n]}_{pp}$
is similar to $f_{ss}$ at short-range with an added long-range tail that
is larger for states of lesser spatial symmetry $[n]$.
Details for these $A=6,7$ trial functions are given in Ref.~\cite{PPCPW97}.

In Type II trial functions, $\phi^{LS}_{p}(r)$ is again the solution of a
$p$-wave differential equation with a potential containing Woods-Saxon and
Coulomb terms, but with an added Lagrange multiplier that turns on at
long range.
This Lagrange multiplier imposes the boundary condition
\begin{equation}
\label{eqn:asymptotic}
[\phi^{LS[n]}_{p}(r\rightarrow\infty)]^n \propto W_{km}(2\gamma r)/r,
\end{equation}
where $W_{km}(2\gamma r)$ is the Whittaker function for bound-state wave
functions in a Coulomb potential and $n$ is the number of $p$-shell nucleons.
The $\gamma$ is related to the cluster separation energy which is taken from
experiment.
The accompanying $f_{sp}$ goes exactly to unity (more rapidly than in the
Type I trial function) and the $f^{LS[n]}_{pp}$ are taken from the exact
deuteron wave function in the case of $^6$Li, or the VMC triton ($^3$He) 
trial function in the case of $^7$Li ($^7$Be).
Consequently, the Type II trial function factorizes at large cluster
separations as
\begin{equation}
\label{eqn:type2}
\Psi_T \rightarrow \psi_{\alpha} \psi_\tau 
                   W_{km}(2\gamma r_{\alpha\tau})/r_{\alpha\tau} \ .
\end{equation}
where $\psi_{\alpha}$ and $\psi_\tau$ are the wave functions of the clusters
and $r_{\alpha\tau}$ is the separation between them.
More details on these wave functions are given in Refs.~\cite{NWS01,N01}.
In the case of $^6$He, which does not have an asymptotic two-cluster
threshold, we generate a $f^{LS[n]}_{pp}$ correlation assuming a weakly bound
$^1$S$_0$ $nn$ pair.

For either type of trial function, a diagonalization is carried out in the 
one-body basis to find the optimal values of the $\beta_{LS[n]}$ mixing
parameters for a given $(J^\pi;T)$ state.
The trial function, Eq.(\ref{eq:psit}), has the advantage of being efficient
to evaluate while including the bulk of the correlation effects.
A more sophisticated variational function can be constructed by including
two-body spin-orbit correlations and additional three-body
correlations~\cite{W91,APW95}, but the time to compute these extra terms is
significant, while the gain in the variational energy is relatively small.
In calculations of $^3$H these extra terms have a negligible ($<0.2\%$)
effect on the GT matrix element.

The wave function at a given spatial configuration
${\bf R}={\bf r}_1,{\bf r}_2,...{\bf r}_A$
can be represented by a vector of $2^A \times I(A,T)$ complex coefficients
in spin and isospin space~\cite{PPCPW97}.
For the nuclei considered here, this gives vectors of 320 in $^6$Li, 576 in
$^6$He, and 1,792 in $^7$Li and $^7$Be.
The spin, isospin, and tensor operators $O^{p=2,6}_{ij}$ contained in the
Hamiltonian and other operators of interest are sparse matrices in this basis.

\section{Nuclear Weak Current}
\label{sec:cur}

The model for the nuclear weak current has been most recently
and exhaustively described in Ref.~\cite{Marcucci01}.  Here
we summarize only its main features.

The nuclear weak current consists of vector and axial-vector parts, 

\begin{eqnarray}
\rho_\pm({\bf q})&=&\rho_\pm({\bf q};V)+\rho_\pm({\bf q};A) \ ,  \\
{\bf j}_\pm({\bf q})&=&{\bf j}_\pm({\bf q};V)+ {\bf j}_\pm({\bf q};A) \ ,
\end{eqnarray}
with corresponding one- and two-body components.  The weak vector current
is constructed from the isovector part of the electromagnetic current,
in accordance with the conserved-vector-current (CVC) hypothesis

\begin{equation}
{\bf j}_{\pm}({\bf q};V)=
 \Big[\, T_{\pm} \, , \, {\bf j}_z({\bf q};\gamma)\, \Big] \ ,
\label{eq:jvv}
\end{equation}
where ${\bf j}_{\pm}({\bf q};V)$ is the charge-lowering (--)
or charge-raising (+) weak vector current,
${\bf j}_z({\bf q};\gamma)$ is
the isovector part of the electromagnetic current, and
$T_{\pm}$ is the (total) isospin-lowering or isospin-raising
operator.  A similar relation holds between the electromagnetic
charge operator and its weak vector counterpart.  For reference,
we list only the expressions for the one-body terms in $\rho({\bf q};V)$
and ${\bf j}({\bf q};V)$, in the notation of Ref.~\cite{Marcucci01}:

\begin{equation}
\rho^{(1)}_i({\bf q};V)= \rho^{(1)}_{i,{\rm NR}}({\bf q};V)+
                       \rho^{(1)}_{i,{\rm RC}}({\bf q};V) \>\>,
\label{eq6}
\end{equation}
with
\begin{equation}
\rho^{(1)}_{i,{\rm NR}}({\bf q};V)= \tau_{i,\pm} \>
 {\rm e}^{{\rm i}{\bf q}\cdot {\bf r}_i} \label{eq7} \ ,
\end{equation}
\begin{equation}
\rho^{(1)}_{i,{\rm RC}}({\bf q};V)=
- {\rm i}\frac{\left ( 2\, \mu^v-1 \right )}{4 m^2} \tau_{i,\pm} \>
{\bf q} \cdot (\bbox{\sigma}_i \times {\bf p}_i) \>
{\rm e}^{ {\rm i} {\bf q} \cdot {\bf r}_i } \ ,
\label{eq8}
\end{equation}
and
\begin{equation}
     {\bf j}^{(1)}_i({\bf q};V)={\frac {1} {2m}} \tau_{i,\pm} \>
   \bigg[ {\bf p}_i\>,\>{\rm e}^{{\rm i} {\bf q} \cdot {\bf r}_i} \bigg]_+
   -{\rm i}  \frac{\mu^v}{2m} \tau_{i,\pm}\>
     {\bf q} \times \bbox{\sigma}_i \> {\rm e}^{{\rm i} {\bf q} \cdot
   {\bf r}_i}  \label{eq9}\>\>\>,
\end{equation}
where $[ \cdots \, ,\, \cdots ]_+$ denotes the anticommutator,
${\bf p}$, $\bbox{\sigma}$, and $\bbox{\tau}$ are
the nucleon's momentum, Pauli spin
and isospin operators, respectively, and $\mu^v$ is
the isovector nucleon magnetic moment ($\mu^v=4.709$ $\mu_N$).  Finally,
the isospin raising and lowering operators are defined as
\begin{equation}
\tau_{i,\pm} \equiv ( \tau_{i,x} \pm {\rm i} \, \tau_{i,y} )/2  \ .
\label{taupm}
\end{equation}

The one-body terms in the axial charge and current
operators have the standard expressions~\cite{Marcucci01}
obtained from the non-relativistic reduction of the covariant
single-nucleon current, and retain terms proportional
to $1/m^2$, $m$ being the nucleon mass:

\begin{equation}
\rho_{i}^{(1)}({\bf q};A)= 
-\frac{g_{A}}{2\,m}\,\tau_{i,\pm}\,{\bbox{\sigma}}
_{i}\cdot\bigg[ {\bf{p}}_{i}\> ,\> 
{\rm e}^{{\rm{i}}{\bf{q}}\cdot{\bf{r}}_{i}}\bigg ]_+ \ ,
\label{eq:rho1}
\end{equation}
and
\begin{equation}
{\bf j}^{(1)}_i({\bf q};A)= {\bf j}^{(1)}_{i,{\rm NR}}({\bf q};A)+
                       {\bf j}^{(1)}_{i,{\rm RC}}({\bf q};A) \ , 
\label{ja1}
\end{equation}
with
\begin{equation}
{\bf j}_{i,{\rm NR}}^{(1)}({\bf q};A)=
-g_{A}\, \tau_{i,\pm} \,{\bbox{\sigma}}_{i}\,
{\rm e}^{{\rm{i}}{\bf{q}}\cdot{\bf{r}}_{i}} \ ,
\label{eq:1baj}
\end{equation}
\begin{eqnarray}
{\bf j}_{i,{\rm RC}}^{(1)}({\bf q};A)=&&
\frac{g_{A}}{4m^2}\,\tau_{i,\pm}\,\Bigg(
{\bbox{\sigma}}_i \bigg[ {\bf p}^2_i\> ,\>
{\rm e}^{{\rm{i}}{\bf{q}}\cdot{\bf{r}}_{i}}\bigg ]_+
-\bigg[ {\bbox{\sigma}}_i \cdot {\bf p}_i \> {\bf{p}}_{i}\> ,\>
{\rm e}^{{\rm{i}}{\bf{q}}\cdot{\bf{r}}_{i}}\bigg ]_+
-\frac{1}{2}{\bbox{\sigma}}_i \cdot {\bf q}  \>\bigg[ {\bf{p}}_{i}\> ,\>
{\rm e}^{{\rm{i}}{\bf{q}}\cdot{\bf{r}}_{i}}\bigg ]_+ \nonumber \\
&& -\frac{1}{2} {\bf q} \> \bigg[ {\bbox{\sigma}}_i \cdot {\bf{p}}_{i}\> ,\>
{\rm e}^{{\rm{i}}{\bf{q}}\cdot{\bf{r}}_{i}}\bigg ]_+
+{\rm i}\> {\bf q} \times {\bf p}_i \>
{\rm e}^{{\rm{i}}{\bf{q}}\cdot{\bf{r}}_{i}} \Bigg)
-\frac{g_P}{2\, m\, m_\mu}\, \tau_{i,\pm} {\bf q} \,
{\bbox \sigma}_i \cdot {\bf q}
\, {\rm e}^{ {\rm i} {\bf q} \cdot {\bf r}_i }\ .
\label{eq:1arc}
\end{eqnarray}

The axial-vector coupling constant $g_A$ is taken
to be~\cite{Adelberger98} 1.2654$\pm$0.0042, by
averaging values obtained from the beta asymmetry
in the decay of polarized neutrons and the half-lives
of the neutron and superallowed $0^+ \rightarrow 0^+$
transitions.  The last term in Eq.~(\ref{eq:1arc}) is
the induced pseudoscalar contribution ($m_\mu$ is the
muon mass), for which the coupling constant $g_P$ is
taken as~\cite{Gmitro87} $g_P$=--6.78 $g_A$.
 
Some of the two-body axial-current operators are derived from
$\pi$- and $\rho$-meson exchanges and the $\rho\pi$-transition
mechanism.  These mesonic operators, first obtained in a systematic way
in Ref.~\cite{Chemtob71}, have been found to give rather small
contributions in weak transitions involving
few-nucleon systems~\cite{Schiavilla98,Marcucci01,Marcucci01a}.
The two-body weak axial-charge operator includes a pion-range term,
which follows from soft-pion theorem and current algebra
arguments~\cite{Kubodera78,Towner92}, and short-range terms,
associated with scalar- and vector-meson exchanges.
The latter are obtained consistently with the two-nucleon
interaction model, following a procedure~\cite{Kirchbach92} similar
to that used to derive the corresponding weak vector-current
operators~\cite{Marcucci01}.

The dominant two-body axial current operator, however,
is that due to $\Delta$-isobar excitation~\cite{Schiavilla98,Marcucci01}.
Since the $N$$\Delta$ transition axial-vector
coupling constant $g_A^*$ is not known experimentally,
the associated contribution suffers from a large model
dependence.  To reduce it~\cite{Schiavilla98,Marcucci01},
the coupling constant $g_A^*$ has been adjusted
to reproduce the experimental value of the Gamow-Teller (GT)
matrix element in tritium $\beta$ decay,
0.957 $\pm$ 0.003~\cite{Schiavilla98}.
The value used for $g_A^*$ in the present work is
$g_A^*$=1.17 $g_A$~\cite{Marcucci01a}. 

The $\Delta$-excitation operator used here is
that derived, in the static $\Delta$ approximation,
using first-order perturbation theory.  This approach
is considerably simpler than that adopted in Ref.~\cite{Marcucci01},
where the $\Delta$ degrees of freedom were
treated non-perturbatively, by retaining them explicitly in
the nuclear wave functions~\cite{Schiavilla92}.  However,
it is important to emphasize~\cite{Marcucci01} that results
obtained within the perturbative and non-perturbative
schemes are within 1\% of each other typically,
once $g_A^*$ is fixed, independently in the perturbative
and non-perturbative calculations, to reproduce
the experimentally known GT matrix element; see
Table~XV in Ref.~\cite{Marcucci01}.
\section{Calculation}
\label{sec:cal}

The calculation of the $\beta$-decay and $\epsilon$-capture
rates proceeds in two steps: firstly, the Monte Carlo
evaluation of the weak charge and current operator
matrix elements, and the subsequent decomposition of
these in terms of reduced matrix elements (RME's);
secondly, the evaluation of the rate by carrying out
the integrations in Eqs.~(\ref{eq:rtt}) and~(\ref{eq:rtte}).

The RME's are obtained from Eqs.~(\ref{eq:c})--(\ref{eq:em})
by choosing appropriately the $\hat{\bf q}$-direction.
For example, in the $\beta$-decay of $^6$He the $C_1(q;A)$
and $L_1(q;A)$ RME's are determined from

\begin{eqnarray}
C_1(q;A)&=&\frac{\rm i}{\sqrt{4\pi}} \langle ^6{\rm Li},10|
 \rho_+^\dagger(q\hat{\bf z};A)|^6{\rm He},00\rangle \ , \\
L_1(q;A)&=&\frac{\rm i}{\sqrt{4\pi}} \langle ^6{\rm Li},10|
 \hat{\bf z}\cdot {\bf j}_+^\dagger(q\hat{\bf z};A) |^6{\rm He},00 \rangle \ ,
\end{eqnarray}
while the $E_1(q;A)$ and $M_1(q;V)$ RME's are determined from

\begin{eqnarray}
E_1(q;A)&=&-\frac{\rm i}{\sqrt{2\pi}} \langle ^6{\rm Li},10|
\hat{\bf z}\cdot {\bf j}_+^\dagger(q\hat{\bf x};A) |^6{\rm He},00 \rangle \ , \\
M_1(q;V)&=&-\frac{1}{\sqrt{2\pi}} \langle ^6{\rm Li},10|
\hat{\bf y}\cdot {\bf j}_+^\dagger(q\hat{\bf x};V) |^6{\rm He},00 \rangle \ .
\label{eq:m1v}
\end{eqnarray}
Here $J_i$,$M_i$=0,0 and $J_f$,$M_f$=1,0 and the
spin-quantization axis is along $\hat{\bf z}$.  The matrix elements
above are computed, without any approximation, by Monte Carlo
integration.  The wave functions are written as
vectors in the spin-isospin space of the $A$ nucleons ($A$=6 or
7 here) for any given spatial configuration
${\bf R}=({\bf r}_1,\dots,{\bf r}_A)$.  For
the given ${\bf R}$, the state vector
$O^\dagger({\bf R}) \Psi_i({\bf R})$ is calculated
with techniques similar to those developed in
Ref.~\cite{Schiavilla89}---$O({\bf R})$ is any of the operators
$\rho({\bf q};V)$, ${\bf j}({\bf q};V)$, etc., and
$\Psi_i({\bf R})$ is the wave function of the initial
nucleus.  The spatial integrations are
carried out with the Monte Carlo method by sampling ${\bf R}$ configurations
according to the Metropolis algorithm, using
a probability density proportional to
$\langle \Psi_f^\dagger({\bf R}) \Psi_f({\bf R})\rangle$,
where $\Psi_f({\bf R})$ is the wave function
of the final nucleus and the notation
$\langle \cdots \rangle$ implies sums
over the spin-isospin states.  Typically 20,000
configurations are enough to achieve a relative
error $\leq$ 1 \% on the matrix elements.

Once the RME's have been obtained, the calculation
of the total transition rate is reduced to performing
the integrations over the outgoing momenta
in Eqs.~(\ref{eq:rtt}) and~(\ref{eq:rtte}).  A glance
at Eqs.~(\ref{eq:rtt}) and~(\ref{eq:rate}) shows that the
differential rate depends on the magnitude of the electron momentum $p_e$
and the variable $x_{e\nu}$=$\hat{\bf p}_e \cdot \hat{\bf p}_\nu$, since
the magnitude of the neutrino momentum is fixed by the
energy-conserving $\delta$-function.  The total rate
can therefore be written as

\begin{equation}
\Gamma^\beta_{fi}= \frac{G_V^2\, m_e^5}{2\, \pi^3} \gamma^\beta_{fi} \ ,
\label{eq:rgfi}
\end{equation}
where the constant $\gamma^\beta_{fi}$ is simply given by

\begin{equation}
\gamma^\beta_{fi}\equiv \frac{2\, \pi}{2\, J_i+1}\,
\int_0^{\overline{p}_e^*} {\rm d}\overline{p}_e\, \overline{p}_e^2 \,
F(Z_f,\overline{p}_e) \int_{-1}^1 {\rm d}x_{e\nu}\, \overline{p}_\nu^2 \,
f_{\rm rec}^{-1} \, \Big[\cdots \Big] \ .
\label{eq:ratet}
\end{equation}
Here all momenta and rest masses have been expressed in units
of the electron mass ($\overline{p}_e$=$p_e/m_e$,
$\overline{m}_i$=$m_i/m_e$, $\overline{m}_e$=1, etc.),
$\Big[\cdots \Big]$ denotes the content of the large
square brackets on the right-hand-side of Eq.~(\ref{eq:rate}),
$f_{\rm rec}^{-1}$ is the recoil factor resulting from
integrating out the $\delta$-function,

\begin{equation}
f_{\rm rec}=\left | 1 + \frac{\overline{p}_e\, x_{e \nu}}{\overline{m}_f}
 + \frac{\overline{p}_\nu}{\overline{m}_f} \right | \ ,
\end{equation}
and the neutrino momentum $\overline{p}_\nu$ is

\begin{equation}
\overline{p}_\nu = \frac{2 \, \overline{\Delta} }
{ 1+\overline{p}_e\, x_{e \nu}/\overline{m}_f
+ \sqrt{ (1+\overline{p}_e \, x_{e \nu}/\overline{m}_f )^2+2\,
\overline{\Delta}/\overline{m}_f } } \ ,
\end{equation}
where $\overline{\Delta}$=$\overline{m}_i-\overline{m}_f
-\sqrt{\overline{p}_e^2+1}-\overline{p}_e^2/(2 \overline{m}_f)$.
Lastly the function $F(Z_f,\overline{p}_e)$ accounts approximately
for wave-function distortion effects of the outgoing electron in the
Coulomb field of the final nucleus with atomic number $Z_f$ and
radius $\overline{R}_f$ (expressed in units of the Compton wavelength of the
electron)~\cite{Schopper66}

\begin{equation}
F(Z,\overline{p}_e)=2\, (1+\gamma_0)\,\,
(2\, \overline{p}_e \overline{R})^{2(\gamma_0-1)} \,\,
\frac{\left|\Gamma(\gamma_0+{\rm i}\nu)\right|^2}
     {\left|\Gamma(2\gamma_0+1)\right|^2}\,\,{\rm e}^{\pi \nu} \ ,
\end{equation}
with $\gamma_0 \equiv \sqrt{1-(\alpha Z)^2}$ and $\nu \equiv \alpha Z/v_e$.
The maximum allowed electron momentum is denoted with $\overline{p}_e^*$,
the upper integration limit in Eq.~(\ref{eq:ratet}), and is given by

\begin{equation}
\overline{p}_e^* = \sqrt{ \left[ \sqrt{\overline{m}_f^2 +2\,
 \overline{m}_f\,(\overline{m}_i-\overline{m}_f)+1}\,-\,\overline{m}_f \right ]^2-1} \ .
\end{equation}
Neglecting the recoil of the final nucleus leads to: $f_{\rm rec}$=1,
$\overline{p}_\nu$=$\overline{m}_i-\overline{m}_f-\sqrt{\overline{p}_e^2+1}$,
and $\overline{p}_e^*$=$\sqrt{\left(\overline{m}_i-\overline{m}_f\right)^2-1}$.

Note that there is an implicit dependence on $p_e$ and $x_{e\nu}$
in the RME's via the momentum transfer $q$.  It is convenient
to make this dependence explicit by expanding the RME's as

\begin{equation}
T_l(q) = q^m\, \sum_{n \ge 0} t_{l,2n} \, q^{2n} \ ,
\end{equation}
consistently with the known expansions of the multipole
operators in powers of $q$~\cite{Marcucci01}.
Here $m=l,l\pm 1$, depending on the RME considered.  For example,
in the $^6$He $\beta$-decay one has $C_1(q;A)$=$q(c_{1,0}+c_{1,2}q^2 +\cdots)$,
$L_1(q;A)$=$l_{1,0}+l_{1,2}q^2 +\cdots$, etc.~.
Given the low momentum transfers involved in all
transitions under consideration, $q \leq 4$ MeV/c, the leading and
next-to-leading order terms $t_{l,0}$ and $t_{l,2}$ are
sufficient to reproduce accurately $T_l(q)$.  Incidentally, the
long-wavelength-approximation corresponds, typically,
to retaining only the $t_{l,0}$ term.  Finally, standard
numerical techniques---Gaussian quadratures---are used
to carry out the integrations in Eq.~(\ref{eq:ratet}). 

It is useful to consider the case of allowed
transitions for which $\left|J_i-J_f\right|$=$\pm 1,0$ and
$\pi_i \, \pi_f$=1, such as the $^6$He $\beta$-decay of
interest in the present study.  Ignoring retardation
corrections due to the finite momentum transfer
involved in the decay, one finds that the only
surviving RME's in the limit $q \rightarrow 0$ are
$C_0(V)$, $L_1(A)$, and $E_1(A)$---of course, $C_0(V)$ vanishes
unless $J_i$=$J_f$.  The one-body terms
in the associated multipole operators read in this
limit~\cite{Marcucci01,Walecka95}:

\begin{eqnarray}
C_{00}(q\!=\!0;V)&=&\frac{1}{\sqrt{4\pi}} \sum_i \tau_{i,\pm} \ , \\
E_{1l_z}(q\!=\!0;A)&=&\sqrt{2}\, L_{1l_z}(q\!=\!0;A)=
-\frac{\rm i}{\sqrt{6 \pi}}\, g_A \sum_i \tau_{i,\pm}\, \sigma_{i,1l_z} \ ,
\end{eqnarray}
where $\sigma_{i,1l_z}$ denote the spherical components of
$\bbox{\sigma}_i$.  The constant $\gamma^\beta_{fi}$ is then
obtained as, neglecting nuclear recoil corrections,

\begin{equation}
\gamma^\beta_{fi}=\frac{|F|^2+g_A^2\,|GT|^2}{2\, J_i+1} \, f
\label{eq:rfgt}
\end{equation}
\begin{equation}
f\equiv \int_1^{\overline{m}_i-\overline{m}_f}
{\rm d}\overline{E}_e\, \overline{E}_e\, \left( \overline{E}_e^2-1\right)^{1/2}\,
\left(\overline{m}_i-\overline{m}_f
-\overline{E}_e\right)^2 \, F(Z_f,\overline{E}_e) \ ,
\label{eq:fff}
\end{equation}
since the integration over $x_{e\nu}$ can now
be performed trivially.  The familiar definitions
of the Fermi and Gamow-Teller (reduced) matrix elements,

\begin{eqnarray}
F&\equiv& \langle J_f||\sum_i\tau_{i,\pm}||J_i\rangle  \ ,
\label{eq:fme} \\
GT&\equiv& \langle J_f||\sum_i\tau_{i,\pm}\, \bbox{\sigma}_i||J_i\rangle  \ , 
\label{eq:gtme}
\end{eqnarray}
have been introduced, as well as $\overline{E}_e$=$\sqrt{\overline{p}_e^2+1}$.
Combining Eqs.~(\ref{eq:rgfi}) and~(\ref{eq:rfgt}) leads
to the standard expression of the decay rate for allowed transitions
as, for example, in Ref.~\cite{Raman78}.

Finally, the total rate for $\epsilon$-capture easily follows
from Eqs.~(\ref{eq:rtte}) and~(\ref{eq:ratee}):

\begin{equation}
\Gamma_{fi}^{\epsilon}=\frac{G_V^2\, m_e^5}{4\, \pi^2}
\, \overline{E}_\nu^2\, |\overline{R}_{1s}(0)|^2\, f^{-1}_{\rm rec} \,
\frac{4\pi}{2\, J_i+1}\,\Big[ \cdots \Big] \ ,
\label{eq:rcap}
\end{equation}
where now $\Big[\cdots \Big]$ denotes the content of the large
square brackets on the right-hand-side of Eq.~(\ref{eq:ratee}),
and the recoil factor is given by
$f_{\rm rec}^{-1}$=$\mid$$1-\overline{E}_\nu/(\overline{m}_i+1)$$\mid\simeq$ 1 with
$\overline{E}_\nu$=$\left[(\overline{m}_i+1)^2-
\overline{m}_f^2\right]/\left[ 2(\overline{m}_i+1)\right]
\simeq \overline{m}_i+1-\overline{m}_f$.  Again, 
we have expressed masses and energies in units of
$m_e$, and $R_{1s}(0)$=$m_e^{3/2} \overline{R}_{1s}(0)$.
In particular, for allowed transitions and ignoring recoil effects, we
obtain the familiar result~\cite{Bambynek77}

\begin{equation}
\Gamma_{fi}^{\epsilon}=\frac{G_V^2\, m_e^5}{4\, \pi^2}
\, \overline{E}_\nu^2\, |\overline{R}_{1s}(0)|^2\, 
\frac{|F|^2+g_A^2\,|GT|^2}{2\, J_i+1} \ .
\label{eq:simple}
\end{equation}

As already pointed out in Sec.~\ref{sec:rate},
however, a more accurate treatment of the atomic physics aspects
is warranted for a meaningful comparison with experiment.  In 
the case of the $^7$Be $\epsilon$-capture of interest
here, such a program has indeed been carried out
by Chen and Crasemann~\cite{Chen78}.  They use multi-configurational
Hartree-Fock (MCHF) wave functions to represent the initial ground state
of the $_4$Be atom as

\begin{equation}
\Psi_i=C_{1}\, \Phi(1s^2\, 2s^2)+C_{2}\, \Phi(1s^2\, 2p^2) \ ,
\end{equation}
and the final $1s$- or $2s$-hole states, after $K$ or $L_I$ captures in
standard nomenclature~\cite{Bambynek77} as

\begin{eqnarray}
\Psi_K&=&C^\prime_1\, \Phi^\prime(1s\, 2s^2)+C^\prime_2\, \Phi^\prime(1s\, 2p^2) \ , \\
\Psi_L&=& \Phi^\prime(1s^2\, 2s) \ . 
\end{eqnarray}
They then proceed to evaluate the matrix elements
$\langle \Psi_K$$\mid$$\psi_e({\bf x}$=$0)$$\mid$$\Psi_i \rangle$
and similarly for $\Psi_L$.  The MCHF approach goes
beyond the independent-particle approximation, commonly
adopted in the analysis of electron capture, by retaining
the effects of Coulomb correlations among the electrons.
The end-result is that the
rate, including both $K$ and $L_I$ captures,
can be conveniently written as in Eq.~(\ref{eq:rcap}),
but for the replacement $|\overline{R}_{1s}(0)|^2\rightarrow
B\times |\overline{R}_{1s}(0)|^2$, where

\begin{equation}
B=B_K\,+\,\Bigg| \frac{\overline{R}_{2s}(0)}{\overline{R}_{1s}(0)}\Bigg|^2\, B_L
\end{equation}
in the notation of Ref.~\cite{Chen78}, with
$B_\alpha$=$\mid$$\langle \Psi_\alpha$$\mid$$\psi_e({\bf x}$=$0)$$\mid$$\Psi_i \rangle$$\mid^2$
and $\alpha$=$K$ or $L$.  Using the values listed in Table~III of Ref.~\cite{Chen78}
for $B_K$, $B_L$, and the ratio of (radial) wave functions at the origin,
as well as the value for $\overline{R}_{1s}(0)$ from Table~IX of
Ref.~\cite{Bambynek77}, we find the relevant combination
$B\times |\overline{R}_{1s}(0)|^2$ to be equal to $7.2403 \times 10^{-5}$
for the case of $^7$Be.
\section{Results}
\label{sec:res}

In this section we report results for the $\beta$-decay of $^6$He
and the $\epsilon$-capture in $^7$Be.  The variational Monte Carlo
(VMC) wave functions of the $A$=6 and 7 nuclei have been obtained
from a realistic Hamiltonian consisting of the Argonne $v_{18}$
two-nucleon~\cite{Wiringa95} and Urbana-IX three-nucleon~\cite{Pudliner95}
interactions, the AV18/UIX model.

The binding energies and radii predicted by the VMC wave functions of Type I 
and II discussed in Sec.~\ref{sec:waves} are listed for reference in 
Table~\ref{tb:be}, along with the exact GFMC energies for this 
Hamiltonian~\cite{PPCPW97} and the experimental values.  
We note that the Type I wave function gives a slightly better energy in 
all cases, except for $^6$Li, while the nucleon radii are very similar in 
all cases except for $^6$He.
The VMC wave functions underbind by about 4 (6) MeV in the $A$=6 (7)
nuclei compared to GFMC; all indications are that this deficiency is due
to a small amount of high-energy ($\sim$1 GeV) contamination,
probably due to inadequate short-range many-body correlations.
In turn, the GFMC energies are about 4\% short of experiment, indicating 
the AV18/UIX model is not quite satisfactory for these nuclei.

The contributions of the different components of
weak vector and axial-vector charge and current operators
to the reduced matrix elements (RME's) contributing
to the $\beta$-decay of $^6$He and the $\epsilon$-capture
in $^7$Be are reported in Tables~\ref{tb:rmes} and~\ref{tb:rmes7e}.
In these tables the column labeled
\lq\lq 1-body\rq\rq~lists the contributions associated
with the one-body terms of the charge and
current operators, including relativistic corrections
proportional to $1/m^2$.  These are the operators
given in Eqs.~(\ref{eq6}),~(\ref{eq9}),~(\ref{eq:rho1}),
and~(\ref{ja1}) of Sec.~\ref{sec:cur}.
The column labeled \lq\lq Mesonic\rq\rq~lists
the contributions from two-body
vector and axial-vector charge and current
operators, associated with pion
and vector-meson exchanges, namely those
of Eqs.~(4.16)--(4.17),~(4.30)--(4.31),~(4.32)--(4.34),
and~(4.35)--(4.37) of Ref.~\cite{Marcucci01}.
Lastly, the column labeled \lq\lq $\Delta$\rq\rq~lists
the contributions arising from $\Delta$
excitation, obtained in perturbation
theory and in the static $\Delta$ approximation,
as in Eqs.~(4.44),~(4.48),~(4.50) and~(4.52) of Ref.~\cite{Marcucci01}.
We reiterate that the coupling constant $g_A^*$ in the $N$$\Delta$
axial two-body current has been set equal to
the value $1.17$ $g_A$, required to reproduce the tritium
Gamow-Teller (GT) matrix element in a calculation based on
on the same treatment of $\Delta$ degrees of freedom, and
using correlated-hyperspherical-harmonics trinucleon wave
functions corresponding to the AV18/UIX Hamiltonian
model~\cite{Marcucci01a} (see discussion in Sec.~\ref{sec:cur}).

The contributions of the different components of
the weak axial current to the GT matrix elements
occurring in the $^6$He $\beta$-decay and
$^7$Be $\epsilon$-capture are reported in
Tables~\ref{tb:l1s},~\ref{tb:l1s7}, and~\ref{tb:l1s7e}.
The notation in these tables is similar to that just
discussed above, with the only difference that it is
now in reference to the axial current only.
Furthermore, the one-body contributions
are separated into the contributions associated
with the leading and next-to-leading terms in the
non-relativistic expansion of the covariant
single-nucleon axial current, Eqs.~(\ref{eq:1baj})
and~(\ref{eq:1arc}).

Having clarified the notation in (most of) the tables,
we now proceed to discuss the results
for the $^6$He $\beta$-decay and $^7$Be $\epsilon$-capture 
separately in the next two subsections.

\subsection{The $^6$He $\beta$-decay}
\label{subsec:res1}

The $^6$He $\beta$-decay is induced by the weak axial-vector
charge and current, and weak vector current operators via the
multipoles $C_1(q;A)$, $L_1(q;A)$, $E_1(q;A)$, and $M_1(q;V)$.
The values for the associated RME's, obtained
with Type I VMC wave functions, are reported in
Table~\ref{tb:rmes} at a value 0.015 fm$^{-1}$ of the lepton
momentum transfer $q$.  Note that in the decay $q$ varies,
ignoring tiny recoil corrections, between 0 and $\simeq$ 0.020
fm$^{-1}$, corresponding to an end-point energy
$m(^6{\rm He})$--$m(^6{\rm Li})$=4.013 MeV/c$^2$.

The largest (in magnitude) RME's are the $L_1(A)$ and
$E_1(A)$ due to transitions induced by the weak axial
current ${\bf j}({\bf q};A)$.  This is to be expected since
the decay $^6$He $\rightarrow$ $^6$Li$\, e^-\, \overline{\nu}_e$
is a superallowed one, having $(J_i^{\pi_i};T_i)$=(0$^+$;1) and
$(J_f^{\pi_f};T_f)$=(1$^+$;0).  The two-body
\lq\lq Mesonic\rq\rq~contributions to ${\bf j}({\bf q};A)$
are individually very small, moreover they interferfere
destructively and nearly cancel out.  The two-body axial
current contributions due to $\Delta$ excitation are
at the level of $\simeq$ 1.5\% of the leading one-body
contributions.  

The transitions induced by the axial charge and vector
current are inhibited, since the first non-vanishing terms
in the long-wavelength expansions of the associated multipole
operators are linear in $q$, and hence their contributions
$C_1(A)$ and $M_1(V)$ suppressed by one power of $qR$
($R$ is the nuclear radius) with respect to $L_1(A)$ and
$E_1(A)$.  The two-body contributions to $M_1(V)$ and
$C_1(A)$ are relatively large, and increase (in magnitude)
the one-body results by 9\% and 25\%, respectively.  Among
the \lq\lq Mesonic\rq\rq~terms, the $\pi$ axial
charge and vector current operators are dominant,
while the vector-meson exchange as well as $\Delta$ operators
play a minor role.  It is important to stress the
model-independent character of the $\pi$-exchange two-body
operators, whose vector and axial-vector structures
are dictated, respectively, by gauge invariance~\cite{Marcucci01}
and current algebra~\cite{Kubodera78} arguments.

The $M_1(V)$ RME is approximately related by CVC to the 
RME of the electromagnetic multipole
operator $M_1(\gamma)$ connecting the $^6$Li$^*$(3.56 MeV) excited state
with $(J^\pi;T)$ assignments (0$^+$;1)
to the $^6$Li(g.s.) ground state $(J^\pi;T)$=(1$^+$;0) via 

\begin{eqnarray}
M_1(q;V)&=&-\frac{1}{\sqrt{2\pi}} \langle ^6{\rm Li},10|
\Big[\, T_{\pm} \, , \, \hat{\bf y}\cdot {\bf j}^\dagger_z(q\hat{\bf x};\gamma)\, \Big] 
|^6{\rm He},00\rangle \nonumber \\
&\simeq&-\sqrt{2} \, M_1(q;\gamma) \ ,
\label{eq:m1gv}
\end{eqnarray}
where we have made use of Eqs.~(\ref{eq:jvv})
and (\ref{eq:m1v}), and have assumed that the
$^6$Li(g.s.) as well as the $^6$He(g.s.) and
$^6$Li$^*$(3.56 MeV) states are members, respectively,
of an isosinglet and an isotriplet.  Of course,
electromagnetic terms and isospin-symmetry-breaking
components in the strong-interaction sector,
both of which are present in the AV18/UIX
Hamiltonian model, will spoil the relation above
to some extent.  However, the present VMC wave functions
of Type I for $^6$Li(g.s.), $^6$Li$^*$(3.56 MeV), and
$^6$He(g.s.) are pure $T=0$ and $TM_T$=10 and 1--1, respectively.
Similar (Type I) wave functions have been recently
employed in Ref.~\cite{Wiringa98} to carry out a calculation
of the elastic and transition form factors of the $A$=6
systems and, in particular, of the radiative width of the
$^6$Li$^*$(3.56 MeV) state.  The values for the
$M_1(\gamma)$ RME were found to be $-{\rm i}\, 2.81 \times 10^{-3}$
and $-{\rm i}\, 3.09 \times 10^{-3}$ including one-body only
and both one- and two-body terms in the electromagnetic
current operator~\cite{Wiringa98} in agreement, on the
basis of Eq.~(\ref{eq:m1gv}), with those reported in Table~\ref{tb:rmes}.
Incidentally, the radiative width of the $^6$Li$^*$(3.56 MeV)
was predicted to be~\cite{Wiringa98} 7.49 eV and 9.06 eV
with one- and (one+two)-body currents.  The experimental
value is ($8.19 \pm 0.17$) eV.

In Table~\ref{tb:l1s} we list results for
the GT RME, related to the $L_1(q=0;A)$ and
$E_1(q=0;A)$ RME's via

\begin{equation}
E_1(q\!=\!0;A)=\sqrt{2}\, L_1(q\!=\!0;A)=
-\frac{\rm i}{\sqrt{6 \pi}}\, g_A\, GT \ ,
\label{eq:e1gt}
\end{equation}
with $g_A$=1.2654.  A few comments are in order.
Firstly, the predicted $^6$He GT RME is about 5\% larger than the value
of 2.173 that is derived from Eqs.~(\ref{eq:rgfi}) and~(\ref{eq:rfgt})
using the most recent tabulation of the log$(ft)$ value
for the $^6$He decay, $2.910 \pm 0.002$, reported
in Ref.~\cite{Tilley01}.  This over-prediction is already
present at the level of the one-body contributions; those
associated with two-body operators further increase
the discrepancy from about 3\% to 5\%.  Secondly,
the difference between the results obtained with Type I and II wave
functions are very small, even though those for $^6$He have very
different nucleon radii.  The Type II wave functions,
in contrast to those of Type I, incorporate long-range
Coulomb correlation effects and the correct two-body
clustering behavior in the asymptotic region, as discussed
in Sec.~\ref{sec:waves}.  However, these asymptotically
improved wave functions have only a marginal impact
on the value of the GT RME, by reducing it by only $\simeq$ 0.2\%.
Thirdly, the relativistic corrections to ${\bf j}({\bf q};A)$
(proportional to $1/m^2$) are relatively large, comparable
to the leading two-body contributions associated with $\Delta$
excitation, and have been neglected in all previous studies
we are aware of.  Lastly, the $E_1(q=0;A)$ RME, derived
from Eq.~(\ref{eq:e1gt}), is $-{\rm i}\,0.6657$ (total),
which should be compared to the value $-{\rm i}\,0.6654$
(total, Type I) obtained at $q$=0.015 fm$^{-1}$.  Thus retardation
corrections are tiny, as expected.

Finally, in Table~\ref{tb:t12} we list the values
for the half-life of $^6$He derived from Eqs.~(\ref{eq:rgfi})
and~(\ref{eq:ratet}) under different approximation
schemes for purpose of illustration (note that
$\tau_{1/2}$=${\rm ln}2/\Gamma$).  The first
row in Table~\ref{tb:t12} is obtained by retaining
only the $L_1(A)$ and $E_1(A)$ RME's evaluated at
$q$=0, namely neglecting retardation corrections
as well as the contributions from transitions
induced by the vector current and
axial charge.  It is equivalent to using Eqs.~(\ref{eq:rfgt})
and~(\ref{eq:fff}), apart from negligible recoiling
corrections.  The second row again includes only the
$L_1(A)$ and $E_1(A)$ RME's, but now keeps their full momentum
transfer dependence. The third row includes all contributing
RME's with their intrinsic $q$-dependence.  The last row
reports the measured half-life from Ref.~\cite{Tilley01}.

It is important to stress that in the present calculation
the effects of Coulomb distortion of the outgoing electron
wave function are considered within the approximate scheme
discussed in Sec.~\ref{sec:cal}.  In existing tabulations,
such as those in Ref.~\cite{Raman78}, these effects
are treated more realistically, by solving the Dirac
equation for the electron in the field generated by the
(extended) charge distribution of the daughter nucleus.
Obviously, this approach complicates considerably the formulae
derived in Sec.~\ref{sec:rate} using plane waves.  However,
the \lq\lq error\rq\rq made in our present treatment
should not be large, as can be inferred from the following
argument.  If one takes the experimental value
for the GT matrix element (2.173) to compute back the
experimental half-life, but using Eqs.~(\ref{eq:rgfi})
and~(\ref{eq:ratet}), one obtains a value of 820.4 ms,
which should be compared to the \lq\lq true\rq\rq~experimental
half-life of 806.7 ms.  This 1.6\% difference is presumably due
to our present approximate scheme for dealing with Coulomb
distortions of the electron waves.  Lastly, if these were to be
altogether ignored [by setting $F(Z,\overline{p}_e)$=1 in
Eq.~(\ref{eq:ratet})], the resulting calculated values for
$\tau_{1/2}$, in the same approximation as in the third
row of Table~\ref{tb:t12}, would be 823.9 ms and 803.1 ms
with 1- and (1+2)-body operators, respectively.

\subsection{The $^7$Be $\epsilon$-capture}
\label{subsec:res2}

The $^7$Be nucleus decays by electron capture to the
ground state of $^7$Li and to its first-excited state
at 0.478 MeV.  The ($J^\pi;T$) assignments of $^7$Be(g.s.),
$^7$Li(g.s.), and $^7$Li$^*$(0.48 MeV) are ($\case{3}{2}^-;\case{1}{2}$), 
($\case{3}{2}^-;\case{1}{2}$),
and ($\case{1}{2}^-;\case{1}{2}$), respectively.

The calculated GT matrix elements for the transitions to the
ground and first excited states of $^7$Li are given in
Tables~\ref{tb:l1s7} and~\ref{tb:l1s7e}.  The first and
second rows in Table~\ref{tb:l1s7} (\ref{tb:l1s7e}) list 
the results obtained with VMC wave functions of Type I, and
based on random walks consisting of 20,000 (15,500) and
10,000 (10,000) configurations, respectively.  Note
that the statistical Monte Carlo errors are at the 1\%
level for the decay to $^7$Li, and almost
an order of magnitude smaller for the decay to
$^7$Li$^*$.  Indeed, the central values computed in
the longer and shorter random walks are fully
consistent (within errors) in the ground- to excited-state
transition, but barely so in the ground- to ground-state
transition.  The third row in
Tables~\ref{tb:l1s7} and~\ref{tb:l1s7e} lists
the results obtained with VMC wave functions of Type II
and based on 10,000 point random walks.  As for the
$A$=6 case, it appears that the asymptotically
improved wave functions of Type II lead to values for the
GT matrix elements not statistically different from those
of Type I.  The two-body contributions increase by roughly
3\% the one-body matrix elements, and as discussed below,
bring theory into better agreement with experiment.

The $\epsilon$-capture to $^7$Li(g.s.) also proceeds
through a Fermi-type transition.  The Fermi matrix
element, defined as in Eq.~(\ref{eq:fme}), is
$F$=$-\sqrt{2\, J_f+1}$ with wave functions of Type I, 
in which isospin-symmetry-breaking components are ignored.
However, the Type II wave functions do include long-range
Coulomb correlations, and therefore break the isospin symmetry
of the isodoublet $^7$Be(g.s.)-$^7$Li(g.s.).  The Fermi matrix
element is calculated to be, in this case, -1.999.

In Table~\ref{tb:rmes7e} we report the contributions from the individual
components of the weak vector and axial charge and current
operators to the dominant RME's occurring in the transition
to the first-excited state of $^7$Li (the corresponding
neutrino energy is $E_\nu$=0.384 MeV or $\overline{E}_\nu$=0.752
in units of the electron mass).  The $M_1(V)$ and $C_1(A)$
transition strengths are down by 3 and 4 order of magnitudes
with respect to the leading $E_1(A)$ and $L_1(A)$.  There are
in principle additional contributions from order 2 multipoles,
such as $C_2(V)$ and $M_2(A)$, for example, however, these
are expected to be even more suppressed than those due to
$M_1(V)$ and $C_1(A)$.  No attempt has been made to calculate them.
One should note that the retardation corrections in
the $E_1(q;A)$ and $L_1(q;A)$ RME's are negligible.  Indeed,
$E_1(q$=$0;A)$=$\sqrt{2}\, L_1(q$=0$;A)$=$6.456 \times 10^{-1}$
in the (1+2)-body calculation, which should be compared with
$6.427 \times 10^{-1}$ from Table~\ref{tb:rmes7e}.  Lastly,
the $M_1(V)$ RME can be expressed via CVC, again ignoring
isospin-symmetry-breaking effects, as 

\begin{equation}
M_1(q;V)\simeq -{\rm i} \frac{\sqrt{2} }{3} \frac{q}{2m}
\Big[ \langle ^7{\rm Be}^*\mid\mid \mu_1(\gamma) \mid\mid ^7{\rm Be}\rangle
     -\langle ^7{\rm Li}^*\mid\mid \mu_1(\gamma) \mid\mid ^7{\rm Li}\rangle \Big ] \ ,
\label{eq:m1ve}
\end{equation}
where $^7$Be$^*$ is the first excited state
of $^7$Be at 0.429 MeV with $(J^\pi;T)$=$(1^-/2;1/2)$, and
$\mu_1(\gamma)$ is the magnetic moment operator.  From the experimentally
known radiative widths of the $^7$Be$^*$ and
$^7$Li$^*$ states~\cite{Tilley01}, the isovector combination above of transition
magnetic moments is found to be ($5.87 \pm 0.14$) $\mu_N$, and
the resulting $M_1(q;V)$ is $-{\rm i}\, 5.66\times 10^{-4}$,
which should be compared to the predicted values
of $-{\rm i}\, 5.010 \times 10^{-4}$ and $-{\rm i}\, 5.904 \times 10^{-4}$
with one- and (one+two)-body currents from
Table~\ref{tb:rmes7e}, respectively.  Incidentally, in the case of the
transition to the ground state of $^7$Li, one finds, using
the experimental values for the $^7$Be and $^7$Li magnetic
moments (respectively, --1.398(15) $\mu_N$ and 3.256424(2) $\mu_N$\ from
Ref.~\cite{Tilley01}), that the isovector combination similar to that
in Eq.~(\ref{eq:m1ve}) (with $^7$Be$^* \rightarrow ^7$Be and
$^7$Li$^* \rightarrow ^7$Li) is
${\rm i}\, (0.1007 \pm 0.0003) \times 10^{-2}$, while the calculated values 
are ${\rm i}\, 0.111 \times 10^{-2}$ and 
${\rm i}\, 0.132 \times 10^{-2}$
with one- and (one+two)-body currents.

Finally, in Table~\ref{tb:tbe} we report the predicted
half-life and branching ratio of $^7$Be to the ground and
first-excited states of $^7$Li.  We have ignored the
tiny corrections due to retardation effects and
transitions induced by the vector current
and axial charge operators, i.e. Eq.~(\ref{eq:simple}) has been
used for $\Gamma^\epsilon$, but for the replacement
$|\overline{R}_{1s}(0)|^2\rightarrow
B\times |\overline{R}_{1s}(0)|^2$=$7.2403 \times 10^{-5}$.
The half-life is over-predicted by about 9\%, while the branching
ratio is under-predicted by 1\%.  Two-body contributions reduce
significantly the discrepancy between the calculated and measured
quantities.
\section{Conclusions}
\label{sec:concl}

In the present study we have reported on calculations of
the $^6$He $\beta$-decay and $^7$Be $\epsilon$-capture rates, using
variational Monte Carlo (VMC) wave functions derived from a realistic
Hamiltonian, and a realistic model for the nuclear weak current
and charge operators, consisting of one- and two-body terms.
Both processes are superallowed, and are therefore driven almost
entirely by the axial current (and, additionally, by the vector charge
in the case of the $^7$Be decay to the ground state of $^7$Li).
The two-body part in the axial current operator
has been adjusted to reproduce the experimentally
known Gamow-Teller (GT) matrix element in $^3$H $\beta$-decay. 

The GT matrix element in $^6$He is over-predicted by about 5\%,
while those in $^7$Be connecting to the ground- and
first-excited states of $^7$Li are under-predicted by 5\%, when compared to
the experimental values.  However, the observed branching ratio in
the $^7$Be $\epsilon$-capture is reasonably well reproduced by
theory.  We have verified explicitly that the (relatively small)
discrepancies between the measured and calculated GT matrix elements
are not explained by the inclusion of retardation effects or by the shift
of strength to suppressed transitions induced by the weak vector
current and axial charge operators. 

One- and two-body axial current contributions interfere constructively,
leading to a 1.7\% (4.4\% on average) increase in the one-body prediction
for the GT matrix element in $^6$He ($^7$Be).  As a result, the inclusion
of two-body operators has the effect of slightly worsening (significantly
improving) the agreement between theory and experiment in
$^6$He ($^7$Be) systems.  It is important
to stress that the same model for the nuclear weak current adopted here,
has been recently shown to provide an excellent description of the
process $^3$He($\mu^- , \nu_\mu$)$^3$H~\cite{Marcucci01a}, in which
two-body (vector and axial-vector) operators contribute 12\% of the total rate.

The origin of the current unsatisfactory situation between theory
and experiment is most likely to be in the approximate character
of the VMC wave functions used here.  We have explored the sensitivity
of the results to alternative VMC wave functions, denoted as Type II 
in Sec.~\ref{sec:waves}, which incorporate a better treatment
of the asymptotic behavior, in particular the clustering properties
into 2+4 or 3+4 sub-systems.  No significant changes in the calculated
values have been found.  Thus, the next logical step in our
quest for a quantitative understanding of weak transitions in the
$A$=6 and 7 (as well as $A$=8) systems, is to repeat the present
calculations with the more accurate Green's function Monte Carlo wave 
functions~\cite{Pudliner95,Wiringa00,Pieper01}.

Another, although probably lesser problem, is the inadequacy of the AV18/UIX
Hamiltonian, as reflected in its underbinding of the $A$=6 and 7 nuclei.
Thus it will also be worthwhile repeating these calculations with one of
the more advanced Illinois three-nucleon potentials~\cite{Pieper01}.
It will also be useful to investigate
the numerical implications of models for the nuclear weak current
derived from effective field theory approaches, such
as those in Refs.~\cite{Park01a,Park01b}.  Most of
the computational techniques and computer codes developed here 
can be carried over to this planned next stage.

Finally, it is important to establish whether
the present approach, based on realistic interactions and currents,
leads to a consistent description of the available experimental
data on weak transitions in light nuclei, {\it beyond} the $A$=3
systems for which its validity has already been ascertained. 
One important implication of such a program should be made clear: it would
serve to corroborate the robustness of our recent predictions for the cross
sections of the proton weak captures on $^1$H~\cite{Schiavilla98,Park01a}
and $^3$He~\cite{Marcucci01,Park01b}.
\section*{Acknowledgments}
The work of R.S.\ was supported by DOE contract DE-AC05-84ER40150
under which the Southeastern Universities Research Association (SURA)
operates the Thomas Jefferson National Accelerator Facility, and that 
of R.B.W.\ by the DOE, Nuclear Physics Division, under contract
No.\ W-31-109-ENG-38.  Most of the calculations were made
possible by grants of computing time from the National Energy Research
Supercomputer Center.
%
%
%

%
%  TABLES
%
%
\begin{table}%[t]
\caption{\label{tb:be}
Binding energies (MeV) and point radii (fm) of $A$=6 and 7 nuclei
obtained with VMC wave functions of Type I and II for the AV18/UIX Hamiltonian
model.  Also listed are the corresponding GFMC and experimental values.}
\begin{tabular}{llddd}
Nucleus         & Wave function & $B$ & $\langle r^2_p \rangle^{1/2}$
                                      & $\langle r^2_n \rangle^{1/2}$    \\
\tableline
$^6$He(g.s.)         & Type I   & 23.99(7) & 1.97(1) & 2.88(2) \\
                     & Type II  & 23.78(7) & 2.28(3) & 3.23(5) \\
                     & GFMC     & 28.1(1)  & 1.97(1) & 2.94(1) \\
                     & Expt     & 29.27    &         &         \\
\tableline
$^6$Li(g.s.)         & Type I   & 27.09(7) & 2.49(1) & 2.49(1) \\
                     & Type II  & 27.34(7) & 2.50(1) & 2.50(1) \\
                     & GFMC     & 31.1(1)  & 2.57(1) & 2.57(1) \\
                     & Expt     & 31.99    & 2.43    &         \\
\tableline
$^7$Be(g.s.)         & Type I   & 30.49(9) & 2.44(1) & 2.30(1) \\
                     & Type II  & 30.00(10)& 2.44(1) & 2.33(1) \\
                     & GFMC     & 36.2(1)  & 2.52(1) & 2.33(1) \\
                     & Expt     & 37.60    &         &         \\
\tableline
$^7$Li(g.s.)         & Type I   & 32.09(9) & 2.30(1) & 2.44(1) \\
                     & Type II  & 31.87(9) & 2.32(1) & 2.43(1) \\
                     & GFMC     & 37.8(1)  & 2.33(1) & 2.52(1) \\
                     & Expt     & 39.24    & 2.27    &         \\
\tableline
$^7$Li$^*$(0.48 MeV) & Type I   & 31.97(9) & 2.29(1) & 2.44(1) \\
                     & Type II  & 31.43(10)& 2.31(1) & 2.42(1) \\
                     & GFMC     & 37.5(2)  &         &         \\
                     & Expt     & 38.76    &         &         \\
\end{tabular}
\end{table}
\begin{table}[bth]
\caption{\label{tb:rmes}
Contributions to the (purely imaginary)
reduced matrix elements $E_1(A)$, $L_1(A)$,
$M_1(V)$, and $C_1(A)$ at $q$=0.015 fm$^{-1}$ in $^6$He
$\beta$-decay.  See text for notation.}
\begin{tabular}{ccccc}
RME      & 1-body   & Mesonic  & $\Delta$ &  (1+2)-body       \\
\tableline
$10^{1}\times E_1(A)$ & --6.540(15)  &--0.004(3)   &--0.110(3) & --6.654(16) \\
$10^{1}\times L_1(A)$ & --4.623(11)  &--0.003(2)   &--0.078(2) & --4.704(11) \\
$10^{3}\times M_1(V)$ &   3.922(9)   &  0.300(2)   &  0.057(1) &   4.279(10) \\
$10^{4}\times C_1(A)$ & --4.584(32)  &--0.977(10)  &--0.171(5) & --5.733(34) \\
\end{tabular}
\end{table}
\begin{table}[bth]
\caption{\label{tb:l1s}
Contributions to the Gamow-Teller matrix element
in $^6$He $\beta$-decay.  See text for notation.}
\begin{tabular}{cccccc}
Wave function & 1-body NR & 1-body RC & Mesonic   & $\Delta$ & (1+2)-body \\
\tableline
Type I   &  2.254(5) & --0.0094(3) & 0.0014(9)  & 0.0376(10) &  2.284(5)    \\
Type II  &  2.246(10)& --0.0100(3) & 0.0011(10) & 0.0418(11) &  2.278(10)   \\
\end{tabular}
\end{table}
\begin{table}[bth]
\caption{\label{tb:t12}
Values in ms for the $^6$He half-life obtained in a number
of approximation schemes.  See text for an explanation.
The measured half-life is also listed.}
\begin{tabular}{ccc}
                          & 1-body & (1+2)-body     \\
\tableline
$E_1(0;A)$ and $L_1(0;A)$ & 762.9  &  743.2         \\
$E_1(q;A)$ and $L_1(q;A)$ & 763.5  &  744.0         \\
All                       & 764.4  &  745.1         \\
Expt                      &        & $806.7\pm 1.5$ \\
\end{tabular}
\end{table}
\begin{table}[bth]
\caption{\label{tb:l1s7}
Contributions to the Gamow-Teller matrix element in
the $^7$Be $\epsilon$-capture to the $^7$Li ground state.
See text for notation.}
\begin{tabular}{cccccc}
Wave function & 1-body NR & 1-body RC & Mesonic   & $\Delta$ & (1+2)-body  \\
\tableline
Type I(20k)  & 2.366(29) & --0.038(2) &  0.0039(18)  & 0.110(3) & 2.441(29) \\
Type I(10k)  & 2.288(42) & --0.034(3) &--0.0029(26)  & 0.110(5) & 2.361(42) \\
Type II(10k) & 2.321(41) & --0.041(2) &--0.0008(9)   & 0.108(5) & 2.387(41) \\
\end{tabular}
\end{table}
\begin{table}[bth]
\caption{\label{tb:l1s7e}
Contributions to the Gamow-Teller matrix element in
the $^7$Be $\epsilon$-capture to the $^7$Li first excited state.
See text for notation.}
\begin{tabular}{cccccc}
Wave function & 1-body NR & 1-body RC & Mesonic   & $\Delta$ & (1+2)-body \\
\tableline
Type I(15.5k) & 2.157(6) & --0.028(1) & 0.0064(8)  & 0.080(1) & 2.215(6)  \\
Type I(10k)   & 2.156(7) & --0.028(1) & 0.0063(10) & 0.080(2) & 2.215(7)  \\
Type II(10k)  & 2.154(7) & --0.034(1) & 0.0049(11) & 0.093(2) & 2.218(8)  \\
\end{tabular}
\end{table}
\begin{table}[bth]
\caption{\label{tb:rmes7e}
Contributions to the (purely imaginary)
reduced matrix elements $E_1(A)$, $L_1(A)$, $M_1(V)$,
and $C_1(A)$ in the $^7$Be $\epsilon$-capture to the $^7$Li first
excited state at $\overline{E}_\nu$=0.7515 (in units of the electron
mass).  See text for notation.}
\begin{tabular}{ccccc}
RME                   & 1-body      & Mesonic     & $\Delta$    &  (1+2)-body  \\
$10^{1}\times E_1(A)$ &  6.192(18)  &  0.016(3)   &  0.220(4)   &  6.427(19)   \\
$10^{1}\times L_1(A)$ &  4.405(13)  &  0.017(2)   &  0.163(3)   &  4.584(13)   \\
$10^{4}\times M_1(V)$ &--5.010(13)  &--0.746(6)   &--0.148(2)   &--5.904(14)   \\
$10^{5}\times C_1(A)$ &  5.328(90)  &  0.456(20)  &  0.163(17)  &  5.948(93)   \\
\tableline
\end{tabular}
\end{table}
\begin{table}[bth]
\caption{\label{tb:tbe} The half-life
and branching ratio of $^7$Be to the ground and first excited states
of $^7$Li, predicted with one- and (one+two)-body
currents, are compared with the experimental values.
The VMC wave functions of Type I are used.}
\begin{tabular}{cccc}
                    & 1-body & (1+2)-body & Expt \\
\tableline
$\tau_{1/2}$(days)  & 62.29  & 58.24      & 53.22 $\pm$ 0.06 \\
$\xi$ (\%)          & 10.20  & 10.33      & 10.44 $\pm$ 0.04 \\
\end{tabular}
\end{table}
\end{document}